\documentclass[a4paper,aip,jap,reprint]{revtex4-1}

\usepackage{graphicx}
\usepackage{amsmath}
\usepackage{bm}

\usepackage{amssymb}

\usepackage{hyperref} 

\usepackage{color}
\usepackage[dvipsnames]{xcolor}

\definecolor{bleu}{rgb}{0,0,1}
\definecolor{rouge}{rgb}{1,0,0}

\definecolor{aclock}{RGB}{162,18,42}
\definecolor{clock}{RGB}{17,143,64}

\newcommand{\Jc}{J_\text{c}}

\usepackage{commath}

\newcommand{\mv}[1]{\mathbf{#1}}

\newcommand{\muv}[1]{\hat{\mv{#1}}}

\newcommand{\mvd}[1]{\dot{\mv{#1}}}

\newcommand{\G}[1]{\mv{G}_{#1}}
\newcommand{\D}[1]{\mv{D}_{#1}}

\newcommand{\Ms}{M_\text{s}}
\newcommand{\NORM}[1]{\left\lVert #1 \right\lVert}

\usepackage{empheq}

\begin{document}

\title[Controlling the synchronization properties of two dipolarly coupled STVOs...]{Controlling the synchronization properties of two dipolarly coupled vortex based spin-torque nano-oscillators by the intermediate of a third one}

\author{Flavio Abreu~Araujo}\email{abreuaraujo.flavio@gmail.com}
\affiliation{Unit\'{e} Mixte de Physique CNRS, Thales, Univ. Paris-Sud, Universit\'{e} Paris-Saclay, F91767 Palaiseau, France}
\altaffiliation[Previous address: ]{Institute of Condensed Matter and Nanosciences, Universit\'{e} catholique de Louvain, BE-1348 Louvainla-Neuve, Belgium}

\author{Julie Grollier}
\affiliation{Unit\'{e} Mixte de Physique CNRS, Thales, Univ. Paris-Sud, Universit\'{e} Paris-Saclay, F91767 Palaiseau, France}

\begin{abstract}
In this paper, we propose to control the strength of phase-locking between two dipolarly coupled vortex based spin-torque nano-oscillators by placing an intermediate oscillator between them. We show through micromagnetic simulations that the strength of phase-locking can be largely tuned by a slight variation of current in the intermediate oscillator. We develop simplified numerical simulations based on analytical expressions of the vortex core trajectories that will be useful for investigating large arrays of densely packed spin-torque oscillators interacting through their stray fields.
\end{abstract}

\maketitle

Spin-torque nano-oscillators are magnetic auto-oscillators of deep submicron dimensions. Made of spin-valves\cite{Kiselev2003, Rippard2004_PRL} or magnetic tunnel junctions,\cite{Deac2008} they can be fabricated on top of a plane of CMOS transistors and they operate at room temperature. The torque on magnetization is generated by sending a spin-polarized current through the ferromagnetic layer. For high enough current densities, this spin-torque can induce sustained magnetization precessions that are then converted into voltage oscillations by magneto-resistive effects. The frequency of these microwave oscillators can be tuned over several GHz by changing the amplitude of the injected dc current or applied magnetic field. Because to this high non-linearity, spin-torque nano-oscillators are sensitive to small variations of magnetic field and electric current\cite{Slavin2009}. In particular, several spin-torque nano-oscillators can mutually synchronize even if their individual frequencies are initially different\cite{Kaka2005, Mancoff2005, Grollier2006, Berkov2013, Sani2013, Siracusano2015, Houshang2016}. Thanks to these features they are excellent candidates for building computing systems inspired from neural synchronization in the brain\cite{Csaba2012_2, Pufall2015, Chen2014_ch15, Yogendra2015}. Indeed bio-inspired computing with oscillators requires to be able to fabricate very large arrays of interacting oscillators, and to be able to control the degree of coupling between the oscillators\cite{Aonishi1998, Hoppensteadt1999}. If several physical phenomena can be used to couple spin-torque oscillators, such as spin waves\cite{Russek2010, Kaka2005, Mancoff2005} or electric currents\cite{Li2010, Grollier2006}, one of the most appealing towards the realization of dense arrays is the dipolar coupling. Indeed when oscillators are closely packed, with edge to edge distance below 500 nm, the dipolar coupling becomes intense and can synchronize their dynamics, as demonstrated theoretically\cite{Belanovsky2012, Belanovsky2013, AbreuAraujo2015} and experimentally\cite{Locatelli2015}. Whereas it is possible to tune the coupling provided by spin waves\cite{Sani2013} and electrical currents\cite{Lebrun2016}, it remains a challenge to modify the interaction originating from the dipolar fields emitted by the oscillators. In this letter, we propose to  adjust the dipolar coupling between two close-by spin-torque oscillators by inserting a third oscillator between them. We study numerically how the amplitude of the current sent through the intermediate oscillator modifies the coupling between the other two. For this purpose, we perform full micromagnetic simulations of the three coupled oscillators in order to have an accurate estimate of the dipolar interactions. Then we develop much faster numerical simulations based on analytical equations for the oscillators' dynamics which will be useful for simulating large scale arrays of dipolarly coupled spin-torque oscillators.

\begin{figure}
	\includegraphics{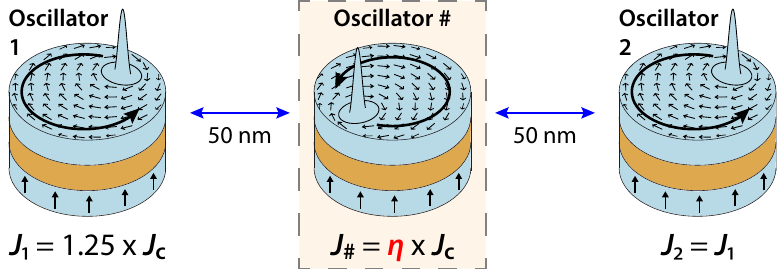}
	\caption{\label{fig:Fig1}Schematic illustration of a three vortex spin-torque oscillators chain where the edge to edge distance is 50 nm. Oscillators 1 \& 2 are supplied with the same current density ($J_1 = J_2$) and the "tuning" oscillator \# is supplied with $J_\text{\#} = \eta \cdot \Jc$.}
\end{figure}

The system we consider is illustrated in Fig. \ref{fig:Fig1}. We study three identical circular nanopillars with diameters $200~\text{nm}$. In each oscillator, the magnetic configuration of the free layer is a vortex, all with the same core polarities and chiralities. We focus on vortex oscillators because the gyration of a vortex core through spin-torque in a single pillar is well understood. It has been shown that analytical descriptions of the dynamics match experiments quantitatively\cite{Dussaux2012}. In addition vortex oscillators have a low phase noise and have been shown to synchronize by dipolar coupling experimentally\cite{Locatelli2015}. In our simulations, we consider that the magnetization of the polarizing magnetic layer is fixed, pointing out of plane, and that the magnetostatic field it emits is negligible. The geometrical and magnetic parameters that we use are displayed in Table \ref{table}.

\begin{table}[ht]
	\begin{tabular}{ccl}
		\hline $h$ &=& 10 nm (dot thickness) \\ 
		$D$ &=& 200 nm (dot diameter) \\ 
		$\Ms$ &=& 800 emu/cm$^3$ (saturation magnetization) \\
		$A$ &=& $1.3 \times 10^{-6}$ erg/cm (stiffness constant) \\ 
		$\alpha$ &=& 0.01 (Gilbert damping parameter) \\
		$P$ &=& 0.2 (current spin polarization) \\
		\hline 
	\end{tabular}
	\caption{\label{table}Geometrical and material parameters for Permalloy (Ni$_{81}$Fe$_{19}$) considered in the simulations.}
\end{table}

In order to study the dynamics of the three dipolarly coupled vortices, we have performed full micromagnetic simulations as well as numerical simulations solving for analytical equations of vortex cores trajectories, and compared both methods.  The micromagnetic simulations are performed using the GPU (Graphics Processing Unit) based micromagnetic code called MuMax$^3$,\cite{mumax3} with a mesh size $2.5 \times 2.5~\text{nm}^2$. The numerical simulations solving for the vortex core gyrotropic motion are based on the Thiele equation \cite{Thiele1973, Guslienko2002, Gaididei2010, Belanovsky2012, AbreuAraujo2015}:
\begin{align}
	\begin{array}{c}
		\displaystyle \G{} \times \mvd{X}_i + \D{} \cdot \mvd{X}_i - \left(k^\text{ms} + k^\text{Oe} J_i\right) \mv{X}_i\\[10pt]
		\displaystyle -\kappa\left( \mv{X}_i \times \muv{z} \right) - \mv{F}^\text{int}_{ij}(\mv{X}_j) = \mv{0}
	\end{array}
	\label{eq:Thiele}
\end{align}
This equation describes the circular motion of the vortex core of position $\mv{X}_i$ in oscillator $i$. The first term is a Magnus-like force, pointing towards the edge of the dot. It arises from the fast upwards spiral of magnetization in the core that generates a gyrovector $\G{i} = -G \muv{z}$.\cite{Thiele1973} The second term accounts for the damping force, tangential to the core trajectory and opposite to the vortex core velocity  $\mvd{X}_i$. The damping coefficient $\D{i}$ is equal to $\D{i} = \alpha  \lambda G (1 + 0.6 s_i^2 - 0.2 s_i^4)$, where $s_i$ is the normalized radius of gyration $s_i = \NORM{\mv{X}_i} / R$ and $\lambda = 0.5 \ln \left( R / (2L_\text{ex}) \right) + 3/8$, with {$L_\text{ex} = \sqrt{A/(2\pi \Ms^2)}$} the exchange length. The third term is the confinement force pointing inwards the dot. It arises both from the magnetostatic energy ($k^\text{ms}_i$) and the current-induced Oersted field confinement ($k^\text{Oe}_i$). The magnetostatic contribution $k^\text{ms}_i$ to the Thiele equation has been calculated under the "Two Vortex Ansatz"\cite{Guslienko2002,Gaididei2010}. We numerically evaluate the energy ($W_\text{ms}$) and specialize our computation to the dot aspect ratio of $\varepsilon = h/ (2R)= 0.05$ and obtain after a polynomial fit:
\begin{multline}
	k_i^\text{ms}(s_i) = \displaystyle \frac{8 \Ms^2 h^2}{R} 1.594 \times\\
	\left( 1 + 0.175 s_i^2 + 0.065 s_i^4 - 0.054 s_i^6 \right).
	\label{eq:kms}
\end{multline}
The $k^\text{Oe}_i$ coefficients are computed using the 10th order Taylor expansion after evaluating the current-induced Oersted field contribution ($W_\text{Oe}$) to the confinement energy and is given by:
\begin{multline}
	k^\text{Oe}_i = J C \Ms h R \frac{8\pi^2}{75} \times\\
	\left( 1 - \frac{4}{7} s_i^2 - \frac{1}{7} s_i^4 - \frac{16}{231} s_i^6 - \frac{125}{3003} s_i^8 + \mathcal{O}\left(s_i^{10}\right) \right).
	\label{eq:kOe}
\end{multline}

The fourth term in Eq. (\ref{eq:Thiele}) is the spin-torque induced force exerted on the vortex core. In our case, since we want to generate sustained gyrations of the core, we choose the current sign so that the spin-torque force points opposite to the damping force. The effective spin torque efficiency is given by $\kappa = \pi a_J M_\text{s} h$ where $a_J$ is the spin torque amplitude $a_J = P \hbar J / (2 |e| M_\text{s} h)$ (with $\hbar$ the Planck constant and $e$ the electron charge). Finally, the last term in Eq. (\ref{eq:Thiele}) accounts for the magnetostatic interaction forces due to stray fields between oscillators $i$ and $j$, the main contribution being dipolar. In contrast to previous works, the analytical version of the magnetostatic interaction developed by Sukhostavets et {\em al.}\cite{Sukhostavets2013} has been considered instead of evaluating it combining micromagnetic simulations and analytical model. $\mv{F}^\text{int}_{ij}$ is given by the following multipole approximation:
\begin{equation}
	\mv{F}^\text{int}_{ij} =  \left[\begin{array}{cc}
		\eta_x^{ij} & 0 \\ 
		0 & \eta_y^{ij}
	\end{array} \right] \mv{X}_j
	\label{eq:Fij}
\end{equation}
where $\eta^{ij}_{x,y}$ are the magnetostatic interaction coefficients:
{\footnotesize
\begin{subequations}
	\begin{empheq}[left=\empheqlbrace]{align}
		\eta^{ij}_{x} &=  \frac{h^2}{R} \Ms^2 \pi^2  \left( \frac{4}{9d_{ij}^3} + \frac{1}{5d_{ij}^5} + \frac{113}{560d_{ij}^7} + \frac{5 \cdot 197}{8 \cdot 16 \cdot 27 d_{ij}^9} \right),\\
		\eta^{ij}_{y} &= - \frac{h^2}{R} \Ms^2 \pi^2  \left( \frac{8}{9d_{ij}^3} + \frac{4}{5d_{ij}^5} + \frac{6 \cdot 113}{560d_{ij}^7} + \frac{5 \cdot 197}{16 \cdot 27 d_{ij}^9} \right),
	\end{empheq}
\end{subequations}}
with $d_{ij} = (2R+L_{ij})/R$ the reduced inter-distance between oscillators ($L_{ij}$ is the edge-to-edge distance between two oscillators).
In this work, the non-linearities of the gyrovector $\G{}$ and the spin-transfer-torque efficiency $\kappa$ have been neglected\cite{Guslienko2014}.

\begin{figure}
	\includegraphics{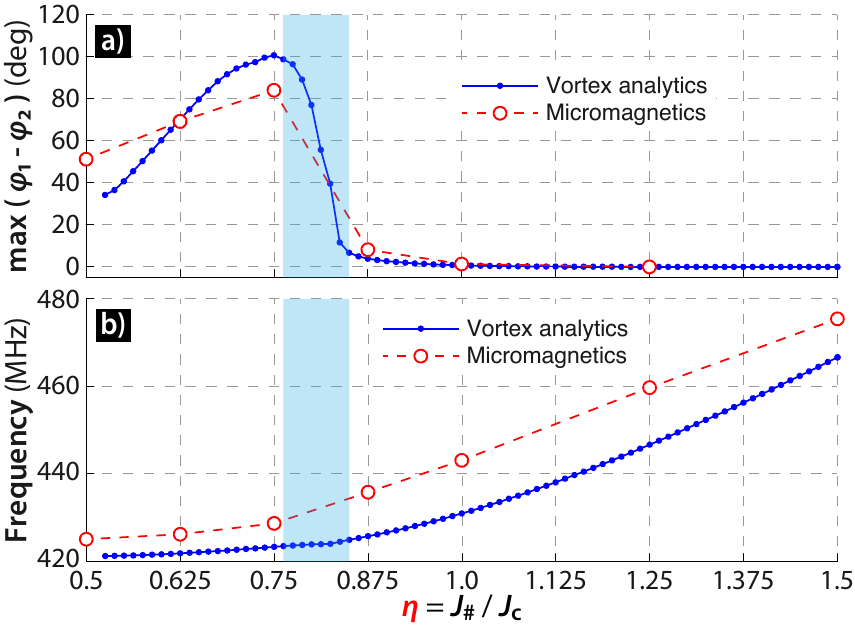}
	\caption{\label{fig:Fig2}(a) Maximum amplitude of the phase difference between oscillators 1 and 2 [$\max(\varphi_2 - \varphi_1)$] (b) Frequency of the oscillators, both as a function of the current through oscillator \#: $J_\# = \eta J_c$. The results of micromagnetic simulations (vortex analytics) are displayed as hollow red circles (small blue disks).}
\end{figure}

The two extreme oscillators labeled 1 and 2 are set in a regime of sustained vortex oscillations by supplying them with a dc current $J_1 = J_2$ above the threshold current for auto-oscillations $\Jc \approx -5.6 \times 10^6~\text{A}/\text{cm}^2$: $J_1 = J_2 = 1.25 \cdot \Jc$. The edge to edge distance between each oscillator is 50 nm, resulting in a separation of 300 nm between oscillators 1 and 2. This distance is small enough for oscillators 1 and 2 to interact strongly through the dipolar fields they emit. In particular, in the absence of the intermediate oscillator \# they mutually synchronize and lock their phases to the same value ($\max(\varphi_2 - \varphi_1) < 2^\circ$))  \cite{Belanovsky2012, AbreuAraujo2015}.  We now study what happens when the intermediate oscillator \# is introduced, by looking at the phase difference $\varphi_2 - \varphi_1$ extracted from micromagnetic and analytically based simulations. Fig. \ref{fig:Fig2}(a) shows the maximum value taken by the phase difference between oscillators 1 and 2, $\max(\varphi_2 - \varphi_1)$, during vortex gyrations as a function of the dc current injected through oscillator \#. 

\begin{figure}
	\includegraphics{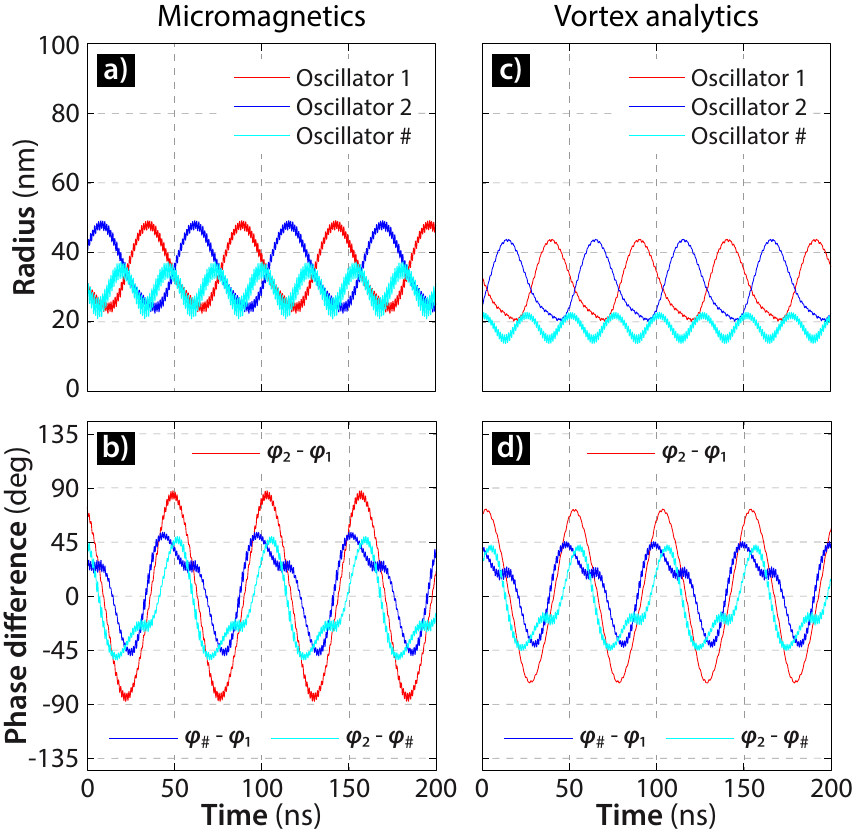}
	\caption{\label{fig:Fig3}(a) and (b) (resp. (c) and (d)) show the radii and phase difference ($\varphi_2 - \varphi_1$) evolutions of the three oscillators illustrated in Fig. \ref{fig:Fig1} for $\eta = J_\# / \Jc = 0.625$ using micromagnetic simulations (resp. vortex analytics).}
\end{figure}

As mentioned in the introduction, the simulations are performed assuming that all the vortex core polarities are parallel. Two regimes appear in Fig. \ref{fig:Fig2}(a). At low currents, for $\eta = J_\# / \Jc < 0.75$, $\max(\varphi_2 - \varphi_1)$ takes large values: the presence of the intermediate vortex destroys the phase locking between oscillators 1 and 2. In the first regime ($0.5 < \eta < 0.86$), modes from the different oscillators can be observed (up to 3) but only the frequency of the main common mode is shown in Fig. \ref{fig:Fig2}(b). Fig. \ref{fig:Fig3} shows time traces of the vortex cores radius and phase differences between each oscillator for $J_\# = 0.625 \cdot \Jc$. Micromagnetic and core-dynamics-based numerical simulations indicate that large fluctuations of the phase differences between the oscillators occur. They also both show that oscillator \# oscillates with a lower amplitude than oscillators 1 and 2. Indeed, in this regime, the current through oscillator \# is much lower than the current $J_c$ leading to auto-oscillations. However, even if the vortex of oscillator \# is damped its orbit fluctuates due to the rotating microwave dipolar fields emitted by oscillators 1 and 2. As $J_\#$ increases the vortex in oscillator \# is less and less damped and its orbit grows. As a result, the dipolar field generated by oscillator \# increases with $J_\#$ and disrupts the trajectories of oscillators 1 and 2 more and more, leading to increasing values of $\max(\varphi_2 - \varphi_1)$ as can be seen in Fig. \ref{fig:Fig2}(a).

\begin{figure}
	\includegraphics{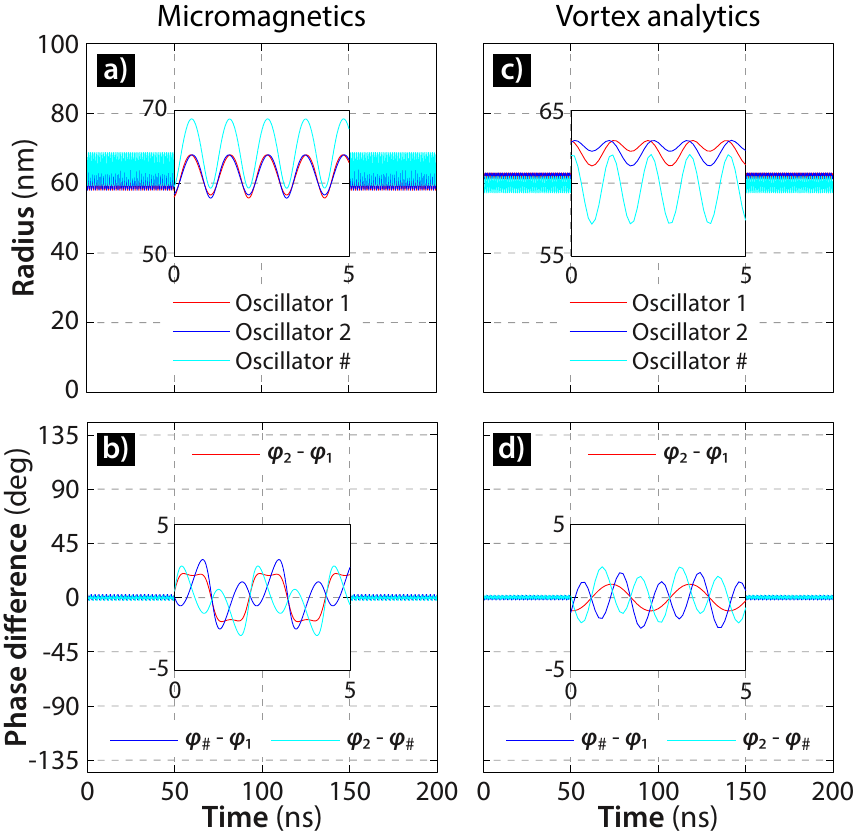}
	\caption{\label{fig:Fig4}(a) and (b) (resp. (c) and (d)) show the radii and phase difference ($\varphi_2 - \varphi_1$) evolutions of the three oscillators illustrated in Fig. \ref{fig:Fig1} for $\eta = J_\# / \Jc = 1.25$ using micromagnetic simulations (resp. vortex analytics).}
\end{figure}

However, for $\eta > 0.86$, a second regime appears where $\max(\varphi_2 - \varphi_1)$ is drastically reduced and phase-locking is restored. In contrast to the first mode, only one common synchronization mode is observed for $\eta > 0.86$. Time traces of the vortex cores radius and phase differences between each oscillator for $J_\# = 1.25 J_c$ are shown in Fig. \ref{fig:Fig4}. Now the radius of all three oscillators is much larger, around 60 nm. Indeed we observe from our simulations that the transition to the phase-locking regime coincides with the onset of self-sustained precessions of oscillator \#. This clearly appears in Fig. \ref{fig:Fig2}(b), which shows how the frequency of  the three coupled oscillators varies with $J_\#$. A kink in the frequency curve appears at the transition to phase-locking. Indeed, while the vortex frequency is practically constant in the damped mode, it increases above the auto-oscillation threshold. In the auto-oscillation regime, the orbit of the vortex core grows with current through spin torque leading to an increase of the confinement and larger frequencies. It should be noted however that the threshold for auto-oscillations of oscillator \# occurs for $\eta <$ 1, in other words below the critical current necessary to compensate the damping. Indeed, the magnetization dynamics of oscillator \# is driven by the dc current $J_\#$ assisted by resonant microwave excitations incoming from oscillators 1 and 2 through their stray fields. As can be seen in Figs. \ref{fig:Fig3} and \ref{fig:Fig3}, small amplitude oscillations of the vortex orbit radii appear. These oscillations are due to a slight shift of the center of the vortex gyrotropic motion (from about 0.1 to about 1 nm depending on the applied current). The frequency of the small amplitude fluctuations is twice the frequency of the main gyrotropic motion.

To summarize, we shown for the first time that the strength of phase-locking between two close-by oscillators interacting via their dipolar fields can be tuned by a slight variation ($0.75 < \eta  < 0.86$ in Fig. \ref{fig:Fig2}(a)) of the current density sent through an intermediate one. In addition, we propose a new full analytical description of the coupled dynamics. Furthermore, numerical simulations based on these analytical expressions of the vortex core dynamics are in excellent agreement with full micromagnetic simulations and several orders of magnitude faster, even with highly efficient solvers making use of GPU hardware. Our results opens the path to the simulation of complex dynamical systems based on large arrays of dipolarly coupled vortex oscillators.

This work is supported by the ANR MEMOS grant (reference: ANR-14-CE26-0021) and idex Nanosaclay. F.A.A. acknowledges the Universit\'e catholique de Louvain for an FSR complement (Fonds Sp\'{e}cial de Recherche).


%

\end{document}